\documentclass[useAMS,usenatbib,usegraphicx]{mn2e}
\bibpunct[,]{(}{)}{;}{a}{}{}

\newcommand{\hm}{h^{-1}}

\title{Statistics of Substructures in Dark Matter Haloes}
\author[E.~Contini et al.]
        {E.~Contini,$^{1,2}$\thanks{Email: contini@oats.inaf.it} 
        G.~De Lucia,$^{2}$
        S.~Borgani,$^{1,2,3}$
        \\      
        $^1$ Dipartimento di Astronomia, Universit{\' a} di Trieste, 
        via G.B. Tiepolo 11, I-34131 Trieste,Italy \\
	$^2$ INAF - Astronomical Observatory of Trieste, 
        via G.B. Tiepolo 11, I-34143 Trieste, Italy \\
	$^3$INFN, Sezione di Trieste, Via Valerio 2, I-34127 Trieste, Italy
        \\}

\pagerange{\pageref{firstpage}--\pageref{lastpage}}
\pubyear{2011}

\begin{document}

\maketitle

\label{firstpage}

\begin{abstract}
We study the amount and distribution of dark matter substructures within dark
matter haloes, using a large set of high-resolution simulations ranging from
group size to cluster size haloes, and carried our within a cosmological model
consistent with WMAP 7-year data. In particular, we study how the measured
properties of subhaloes vary as a function of the parent halo mass, the
physical properties of the parent halo, and redshift. The fraction of halo mass
in substructures increases with increasing mass: it is of the order of 5 per
cent for haloes with ${\rm M}_{200}\sim 10^{13}\,{\rm M}_{\odot}$ and of the
order of 10 per cent for the most massive haloes in our sample, with ${\rm
  M}_{200}\sim 10^{15}\,{\rm M}_{\odot}$. There is, however, a very large
halo-to-halo scatter that can be explained only in part by a range of halo
physical properties, e.g. concentration. At given halo mass, less concentrated
haloes contain significantly larger fractions of mass in substructures because
of the reduced strength of tidal disruption. Most of the substructure mass is
located at the outskirts of the parent haloes, in relatively few massive
subhaloes. This mass segregation appears to become stronger at increasing
redshift, and should reflect into a more significant mass segregation of the
galaxy population at different cosmic epochs. When haloes are accreted onto
larger structures, their mass is significantly reduced by tidal
stripping. Haloes that are more massive at the time of accretion (these should
host more luminous galaxies) are brought closer to the centre on shorter
time-scales by dynamical friction, and therefore suffer of a more significant
stripping. The halo merger rate depends strongly on the environment with
substructure in more massive haloes suffering more important mergers than their
counterparts residing in less massive systems. This should translate into a
different morphological mix for haloes of different mass.
\end{abstract}


\section[]{Introduction}
\label{sec:intro}

In the currently accepted $\Lambda$CDM paradigm for cosmic structure formation,
small dark matter haloes form first while more massive haloes form later
through accretion of diffuse matter and mergers between smaller systems. During
the last decades, we have witnessed a rapid development of numerical
algorithms and a significant increase in numerical resolution, that have
allowed us to improve our knowledge of the formation and evolution of dark
matter structures. In particular, the increase in numerical resolution has
allowed us to overcome the so-called {\it overmerging problem}, i.e. the rapid
disruption of galaxy-size substructures in groups and clusters \citep[][ and
  references therein]{klypin}. If any, we are now facing the opposite problem,
at least on galaxy scales, where many more substructures than visible dwarf
galaxies are found \citep[][ and references therein]{ishiyama,tikhonov}. 

According to the two stage theory proposed by \citet{white}, the physical
properties of galaxies are determined by cooling and condensation of gas within
the potential wells of dark matter haloes. Therefore, substructures represent
the birth-sites of luminous galaxies, and the analysis of their mass and
spatial distribution, as well as of their merger and mass accretion histories
provide important information about the expected properties of galaxies in the
framework of hierarchical galaxy formation models.

Nowadays, a wealth of substructures are routinely identified in dissipationless
simulations, and their statistical properties and evolution have been studied
in detail in the past years. The identification of dark matter substructures,
or {\it subhaloes}, remains a difficult technical task that can be achieved
using different algorithms \citep[see e.g.][]{knebe}. Each of these has its own
advantages and weaknesses, and different criteria for defining the boundaries
and membership of substructures are likely leading to systematic differences
between the physical properties of subhaloes identified through different
algorithms.  However, these might be probably corrected using simple scaling
factors, as suggested by the fact that different studies find very similar
slopes for the subhalo mass function, i.e. the distribution of substructures as
a function of their mass. This is one of the most accurately studied properties
of dark matter substructures, although it remains unclear if and how it depends
on the parent halo mass. \citet{moore} used one high-resolution simulation of a
cluster-size halo and one high-resolution simulation of a galaxy-size halo, and
found that the latter can be viewed as a scaled version of the former. Later
work by \cite{delucia} used larger samples of simulated haloes, but found no
clear variation of the subhalo mass function as a function of the parent halo
mass. Such a dependency was later found by \citet{gao} and \citet{gao2}, who
showed that the subhalo mass function varies systematically as a function of
halo mass and halo physical properties like concentration and formation time.

Typically, only about 10 per cent of the total mass of a dark matter halo is
found in substructures. In addition, their spatial distribution is found to be
{\it anti-biased} with respect to that of dark matter
\citep{ghigna2,delucia,nagai,saro}. It is unclear if the radial distribution of
substructures depends on the parent halo mass. \citet{delucia} found hints for
a steeper radial number density profiles of substructures in low mass haloes
than in high mass haloes. They used, however, a relatively small sample of
simulated haloes, that were run with different codes and numerical
parameters. In this study, we will re-address this issue by using a much larger
sample of simulated haloes, all run with the same code and numerical parameters.

Most previous work focusing on dark matter substructures has studied their
properties as a function of their {\it present day} mass. This quantity cannot
be, however, simply related to the luminosity of the galaxies residing in the
substructures under consideration. Indeed, dark matter substructures are very
fragile systems that are strongly affected by tidal stripping
\citep{delucia,gao}. Since this process affects primarily the outer regions of
subhaloes, and galaxies reside in their inner regions, it is to be expected
that the galaxy luminosity/stellar mass is more strongly related to the mass of
the substructure at the time of {\it infall} (i.e. before becoming a
substructure) than at present \citep{gao3,wang,vale}. In this paper, we will
study the evolution of dark matter substructures splitting our samples
according to different values of the mass at infall.

In this paper, we take advantage of a large set of N-body simulations covering
a wide dynamical range in halo mass, and with relatively high-resolution. This
will allow us to study how the statistical properties of substructures vary as
a function of halo mass, cosmic epoch, and physical properties of the parent
halo. The layout of the paper is as follows: in section~\ref{sec:simulations},
we introduce the simulation set and samples used in our study. In
section~\ref{sec:properties}, we study how the subhalo mass function and
subhalo spatial distribution vary as a function of halo mass, redshift and
concentration. In the second part of our paper (section \ref{sec:subevol}), we
discuss the mass accretion and merging histories of subhaloes as a function of
their mass, accretion time, and environment. Finally, in Section
\ref{sec:concl}, we discuss our findings and give our conclusions.

\section[]{Cluster simulations}
\label{sec:simulations}

Our set of DM haloes is based on `zoom-in' simulations of 27 Lagrangian regions
extracted around massive dark matter haloes, originally identified within a
low-resolution N-body cosmological simulation. For a detailed discussion of
this simulation set, we refer to \citet[][ see also
  \citealt{fabjan}]{bonafede}.  The parent simulation followed $1024^3$ DM
particles within of a box of $1\, h^{-1}$Gpc comoving on a side. The adopted
cosmological model assumed $\Omega_m=0.24$ for the matter density parameter,
$\Omega_{\rm bar}=0.04$ for the contribution of baryons, $H_0=72\,{\rm
  km\,s^{-1}Mpc^{-1}}$ for the present-day Hubble constant, $n_s=0.96$ for the
primordial spectral index, and $\sigma_8=0.8$ for the normalization of the
power spectrum. The latter is expressed as the r.m.s. fluctuation level at
$z=0$ within a top-hat sphere of $8\,\hm$Mpc radius. With this parameters
choice, the assumed cosmogony is consistent with constraints derived from
seven-year data from the Wilkinson Microwave Anisotropy Probe (WMAP7,
\citealt{komatsu}).

The selected Lagrangian regions were chosen so that 13 of them are centred
around the 13 most massive clusters found in the cosmological volume, all
having virial\footnote{ Here we define the virial mass ($M_{\rm 200}$) as the
  mass contained within the radius $R_{\rm 200}$, that encloses a mean density
  of 200 times the critical density of the Universe at the redshift of
  interest.}  mass $M_{\rm 200} \simeq 10^{15}{\,\hm M}_{\odot}$. Additional
regions were chosen around clusters in the mass range $M_{\rm 200}\simeq
(5-10)\times 10^{14}{\,\hm M}_{\odot}$. Within each Lagrangian region, we
increased mass resolution and added the relevant high-frequency modes of the
power spectrum, using the Zoomed Initial Condition (ZIC) technique presented by
\cite*{tormen}.  Outside the regions of high--resolution, particles of mass
increasing with distance are used, so that the computational effort is
concentrated on the cluster of interest, while a correct description of the
large--scale tidal field is preserved. For the simulations used in this study,
the initial conditions have been generated using $m_{\rm DM} = 10^8 {\,\hm
  M}_{\odot}$ for DM particle mass in the high--resolution regions. This mass
resolution is a factor 10 better than the value used by
\cite{bonafede} and \citet{fabjan} to carry out hydrodynamic simulations for
the same set of haloes.

\begin{table}
\caption{Our simulation set has been split in five subsamples, according to the
  halo mass. In the first column, we give the name of the subsample, while the
  second column indicates the range of $M_{200}$ values corresponding to
    each subsample. The third and fourth columns give the number of haloes and
  mean number of subhaloes (with mass above $2\cdot10^9 \,\hm M_{\odot}$)
  within the virial radius ($R_{200}$), respectively.}
\begin{center}
\begin{tabular}{llllll}
\hline
Name & Mass range & $N_{haloes}$ & $ \bar{N}_{subs} $\\
\hline

S$1$ & $ \geq 10^{15}\,\hm {\rm M}_{\odot}$ & 13 & 2943 \\

S$2$ & $[$5-10$]\times10^{14}\,\hm {\rm M}_{\odot}$ &  15 & 1693 \\ 

S$3$ & $[$1-5$]\times10^{14}\,\hm {\rm M}_{\odot}$ & 25 &  358 \\ 

S$4$ & $[$5-10$]\times10^{13}\,\hm {\rm M}_{\odot}$ & 29 & 146 \\ 

S$5$ & $[$1-5$]\times10^{13}\,\hm {\rm M}_{\odot}$ & 259 & 40 \\
\hline
\end{tabular}
\end{center}
\label{tab:tab1}
\end{table}

Using an iterative procedure, we have shaped each high--resolution
Lagrangian region so that no low--resolution particle `contaminates'
the central `zoomed in' halo, out to 5 virial radii of the main
cluster at $z=0$. In our simulations, each high resolution region is
sufficiently large to contain more than one interesting massive halo,
with no `contaminants', out to at least one virial radius. Our
  final sample contains 341 haloes with mass larger than $10^{13} \hm
  M_{\odot}$. We have split this sample into 5 different subsamples,
  as indicated in Table~\ref{tab:tab1}, where we list the number of
  non-contaminated haloes for each sample and the mean number of
  substructures per halo in each subsample.

Simulations have been carried out using the Tree-PM GADGET-3 code. We
adopted a Plummer--equivalent softening length for the computation of
the gravitational force in the high--resolution region. This is fixed
to $\epsilon=2.3\,h^{-1}$~kpc in physical units at redshift $z<2$, and
in comoving units at higher redshift. For each simulation, data have
been stored at $93$ output times between $z\sim 60$ and $z=0$. Dark
matter haloes have been identified using a standard friends-of-friends
(FOF) algorithm, with a linking length of 0.16 in units of the mean
inter-particle separation in the high-resolution region. The algorithm
{\small SUBFIND} \citep{springel} has then been used to decompose each
FOF group into a set of disjoint substructures, identified as locally
overdense regions in the density field of the background halo. As in
previous work, only substructures which retain at least 20 bound
particles after a gravitational unbinding procedure are considered to
be genuine substructures. Given our numerical resolution, the smallest
structure we can resolve has a mass of $M=2\times 10^9 \,\hm
M_{\odot}$. To avoid being too close to the resolution limit of the
simulations, we will sometimes consider only substructures that
contain at least 100 particles, i.e. we will adopt a mass limit of
$1\times10^{10}\,\hm M_{\odot}$.

\section[]{Amount and distributions of dark matter substructures}
\label{sec:properties}

In this section we will consider some basic statistics of the dark matter
substructures in our sample. In particular, we will address the following
questions: what is the mass fraction in substructures? What is their mass and
spatial distribution? And how do these properties vary as a function of the
halo mass, or as a function of other physical properties of the parent haloes?
As discussed above, if subhaloes are to be considered the places where galaxies
are located, these statistics provide us important information about the
statistical properties of cluster galaxy populations expected in hierarchical
cosmologies. 

\subsection[]{Mass Fraction in Subhaloes}
\label{sec:massfrac}

Previous work has found that only 5 to 10 per cent of the halo mass is
contained in substructures, with most of it actually contained in relatively
few massive subhaloes \citep{ghigna,ghigna2,springel,stoehr,gao,delucia}.

\begin{center}
\begin{figure*}
\begin{tabular}{c}
\includegraphics[scale=.65]{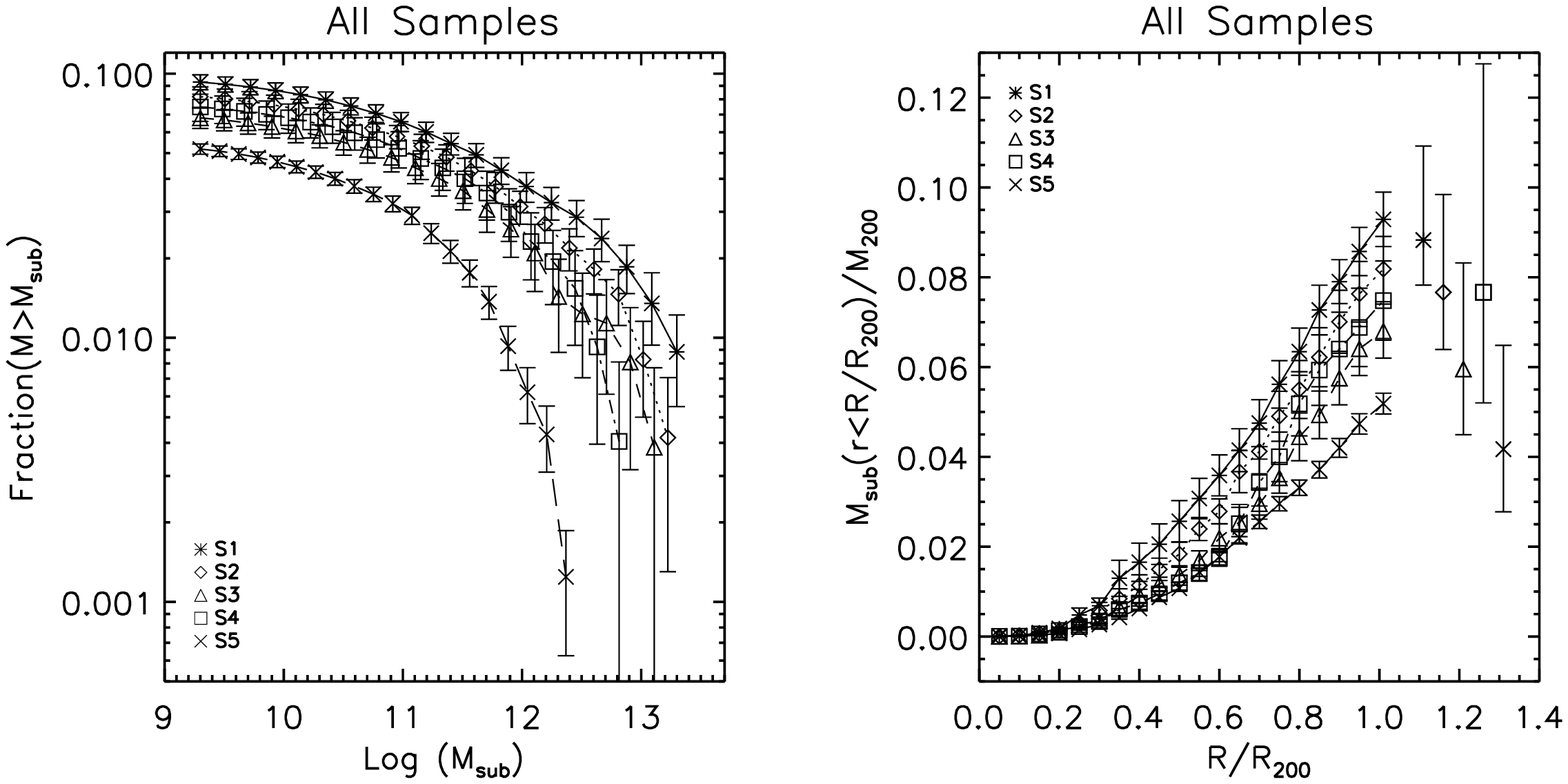} \\
\includegraphics[scale=.65]{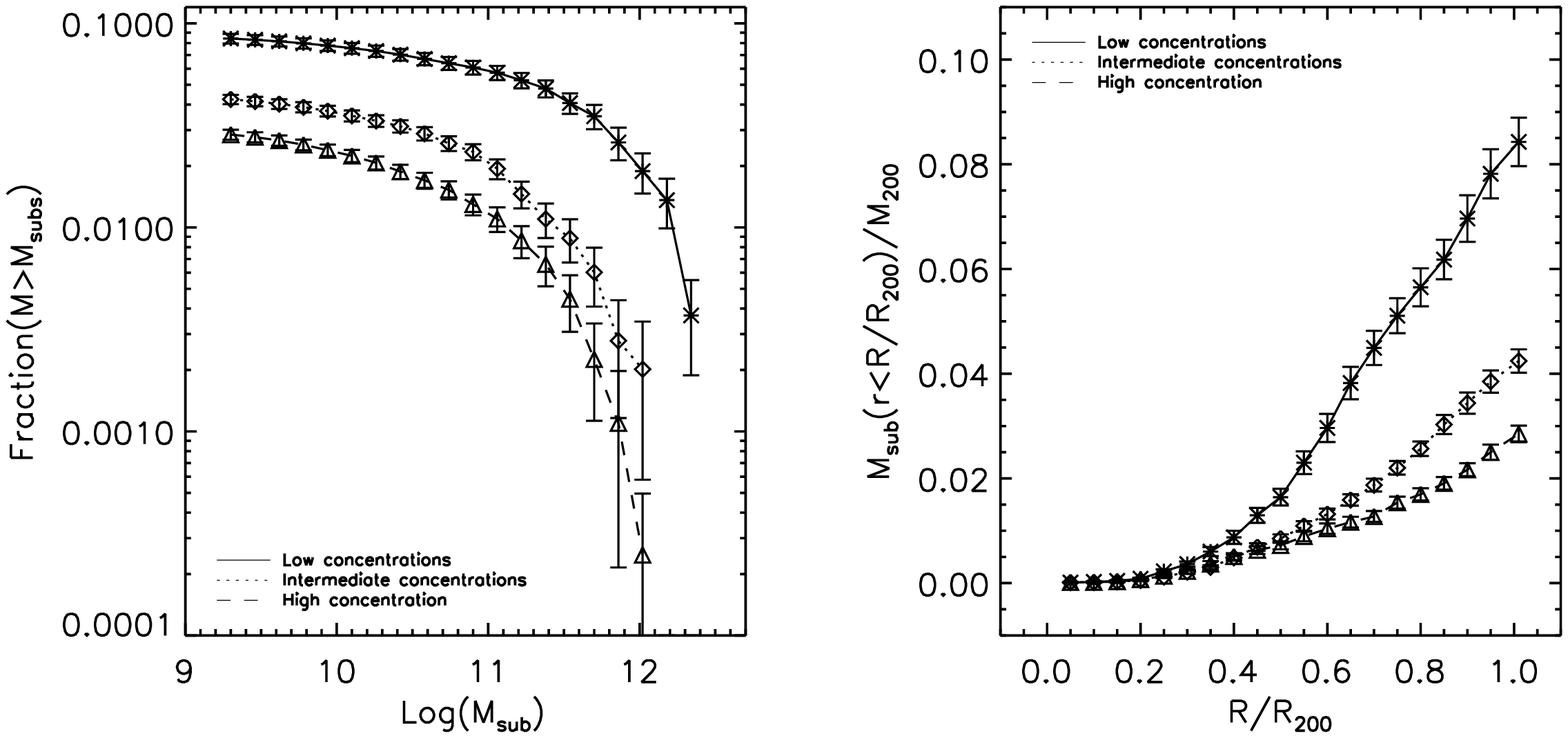} \\
\end{tabular}
\caption{Top panels: cumulative mass fraction in substructures as a function of
  subhalo mass (left) and normalized distance from the halo centre (right), for
  the five samples used in this study (different symbols, as indicated in the
  legend). In the right panel the rightmost symbols with error bars show the
  median, $25^{th}$ and $75^{th}$ percentile of the distributions at
  $R/R_{200}=1$. Bottom panels: same as in the top panels but using only haloes
  from our sample S5 (the least massive one), and splitting the sample in three
  different bins according to the concentration of the parent haloes. In all
  panels, symbols connected by lines show the mean values, while error bars
  show the rms scatter around the mean.}
\label{fig5}
\end{figure*}
\end{center}

Results for our simulation set are shown in Figure~\ref{fig5}. The top left
panel shows the cumulative mass fraction in subhaloes above the mass indicated
on the x-axis, for the five samples considered in our study. There is a clear
trend for an increasing mass in substructures for more massive haloes. For our
most massive sample (S1), about ten per cent of the halo mass is contained in
substructures more massive than $2\times 10^9 \,\hm M_{\odot}$, and
approximately ten per cent of the mass in substructures is contained in the
most massive ones. For less massive haloes, the mass fraction in substructures
decreases. 

Most of the substructures are located outside the central core of dark matter
haloes. In particular, the top right panel of Figure~\ref{fig5} shows that the
substructure mass fraction is smaller than $\sim 1$ per cent out to $\sim
0.3\times r_{200}$, and increases to half its total (within $r_{200}$) value at
$\sim 0.8\times r_{200}$. The results shown can be explained by considering
that haloes of larger mass are less concentrated and dynamically younger than
their less massive counterparts. As we will show below, and as discussed in
previous studies, subhaloes are strongly affected by dynamical friction and
tidal stripping. Less massive haloes assemble earlier than their more massive
counterparts, i.e.  accrete most of the haloes that contribute to their final
mass at early times, so that there was enough time to `erase' the structures
below the resolution of the simulation in these systems. In addition, haloes
that were accreted earlier, and therefore suffered of tidal stripping for
longer times, are preferentially located closer to the centre \citep[see Figure
  15 in][]{gao}.

  For haloes of the same mass, a relatively large range of
  concentrations is possible so that a range of mass fractions is
  expected.  This is confirmed in the bottom panels of
  Figure~\ref{fig5} where we have considered only haloes in our least
  massive sample (S5), and split it into three different bins
  according to the halo concentration so as to have the same number of
  haloes for each bin. We approximate the concentration by
    $V_{max}/V_{200}$, where $V_{max}$ is the maximum circular
    velocity, which is computed by considering all particles bound to
    a given halo, while
    $V_{200}=\sqrt{GM_{200}/R_{200}}$. Interestingly, the lowest
  concentration bin contains substructure mass fractions that are, on
  average, very close to those of our most massive samples (S1 in the
  top panels). This confirms that the halo to halo scatter is very
  large, and that it can be explained only in part by haloes in the
  same mass bin covering a range of physical properties. In order to
  give an idea of the intrinsic scatter of haloes in the same mass
  bin, we have repeated the last point in the top right panel of
  Figure~\ref{fig5}, showing this time the median and the 25th and
  75th percentile of the distributions obtained at $R/R_{200}=1$.

\subsection[]{Subhalo Mass Function}
\label{sec:submf}

One of the most basic statistics of the subhalo population is provided by the
subhalo mass function, i.e. the distribution of dark matter substructures as a
function of their mass. This has been analysed in many previous studies with
the aim to answer the following questions: does the subhalo mass function vary
as a function of the parent halo mass? How does it vary as a function of cosmic
time? And as a function of halo properties (e.g. concentration, formation time,
etc.)?

\begin{figure}
\begin{center}
\includegraphics[scale=.50]{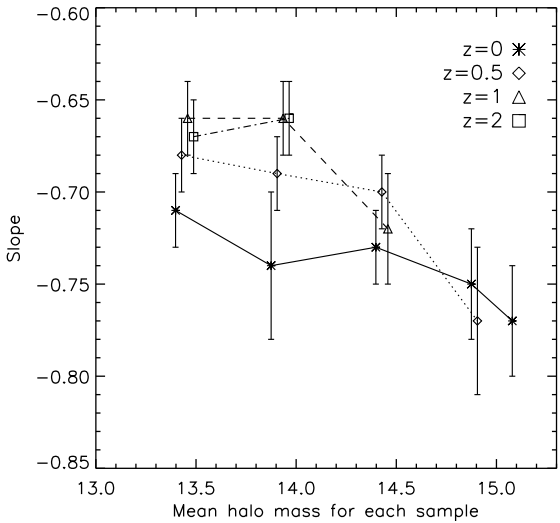}
\caption{Slope of the differential mass function measured for the
    different samples considered in this study, at different cosmic
    epochs (solid line for $z=0$, dotted for $z=0.5$, dashed line for
    $z=1$, and dash-dotted line for $z=2$). Error bars are computed as
    the standard deviation of the slopes measured for each halo within
    the sample. For reasons of clarity, a small shift has been added
    to the abscissa.}
\label{fig2}
\end{center}
\end{figure}

First studies were based on very small samples of simulated haloes,
and claimed the `universality' of the subhalo mass
function. E.g. \citet{moore} compared the substructure mass
distribution obtained for one simulated cluster of mass similar to
that of the Virgo Cluster, and one simulated galaxy-size halo, and
argued that galactic haloes can be considered as `scaled versions' of
cluster-size haloes. \citet{delucia} used a sample of $\sim 11$ high
resolution resimulations of galaxy clusters together with a simulation
of a region with average density. They argued that the subhalo mass
function depends {\it at most weakly} on the parent halo mass and that
the (nearly) invariance of the subhalo mass function could lie in the
physical nature of the dynamical balance between two opposite effects:
the destruction of substructures due to dynamical friction and tidal
stripping on the one hand, and the accretion of new substructures on
the other hand. Contemporary work by \citet[][]{gao} and later work
\citep[e.g.][]{gao2} has demonstrated that the subhalo mass function
does depend on the parent halo mass, as well as on the physical
properties of the parent halo, in particular its concentration and
formation time. We note that \citet{gao} used a sample of
  simulated haloes that was not homogeneous in terms of resolution
  (typically lower than ours), cosmological parameters, and simulation
  codes. The sample used in \citet{gao2} was instead based on a ho
  homogeneous set of cosmological parameters (consistent with WMAP
  first-year results) and included simulations with resolution higher
  than that of our sample. Their sample, however, did not include very
  massive haloes ($\sim 10^{15}\,\hm M_{\odot}$). It is therefore
  interesting to re-address the questions listed above using our
  simulation sample.

\begin{center}
\begin{figure*}
\begin{tabular}{cc}
\includegraphics[scale=0.50]{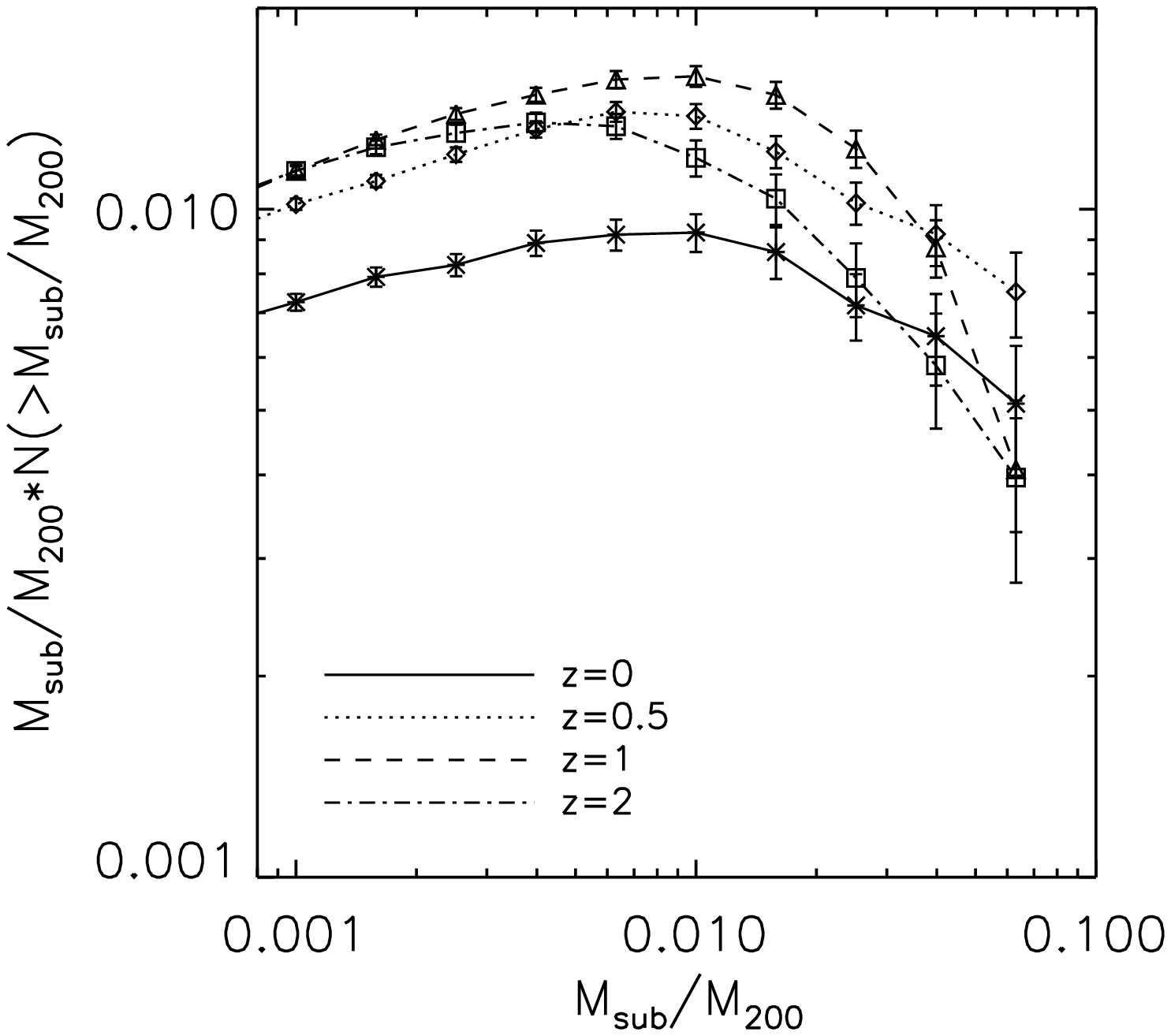} &
\includegraphics[scale=0.50]{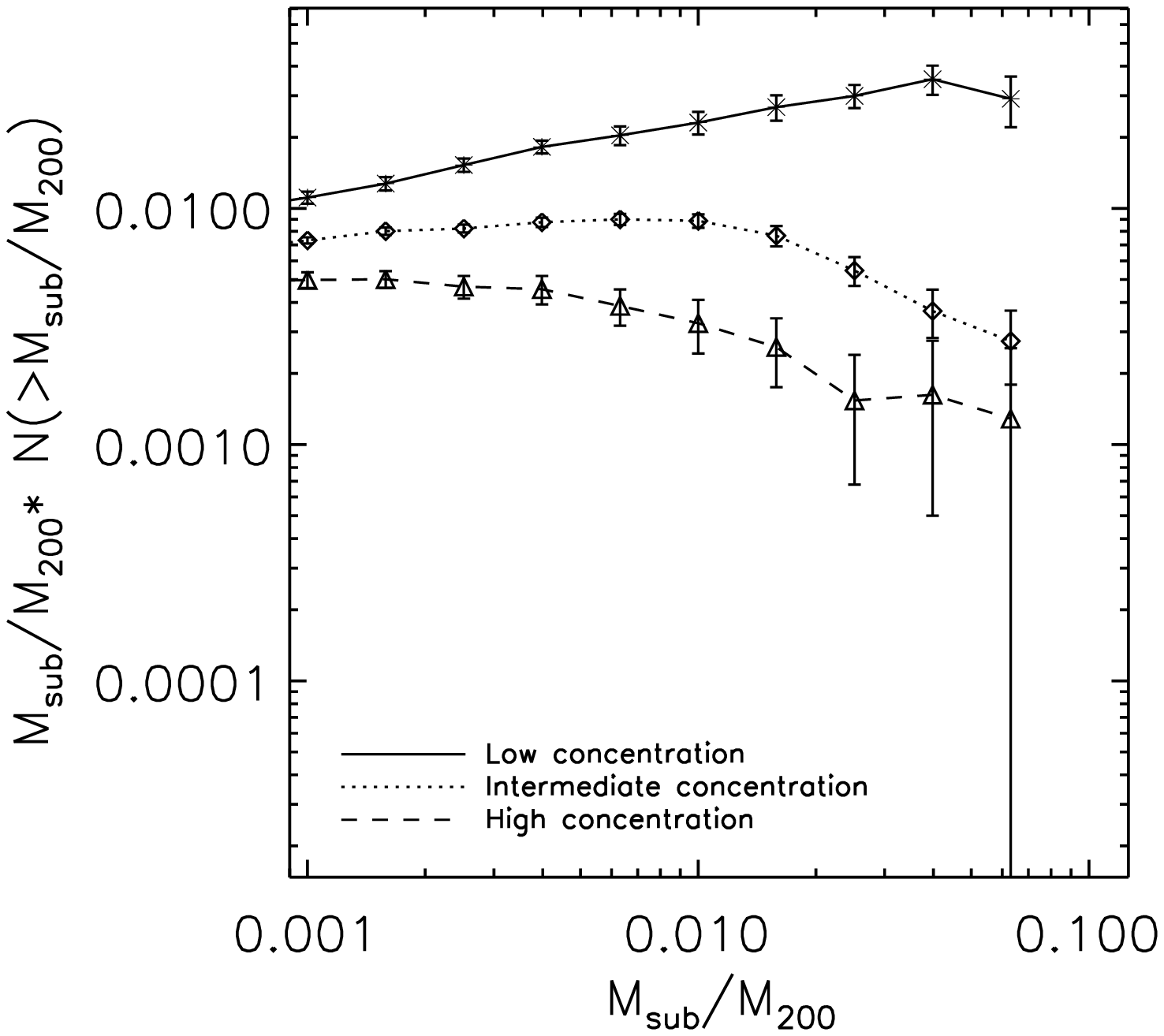} \\
\includegraphics[scale=0.50]{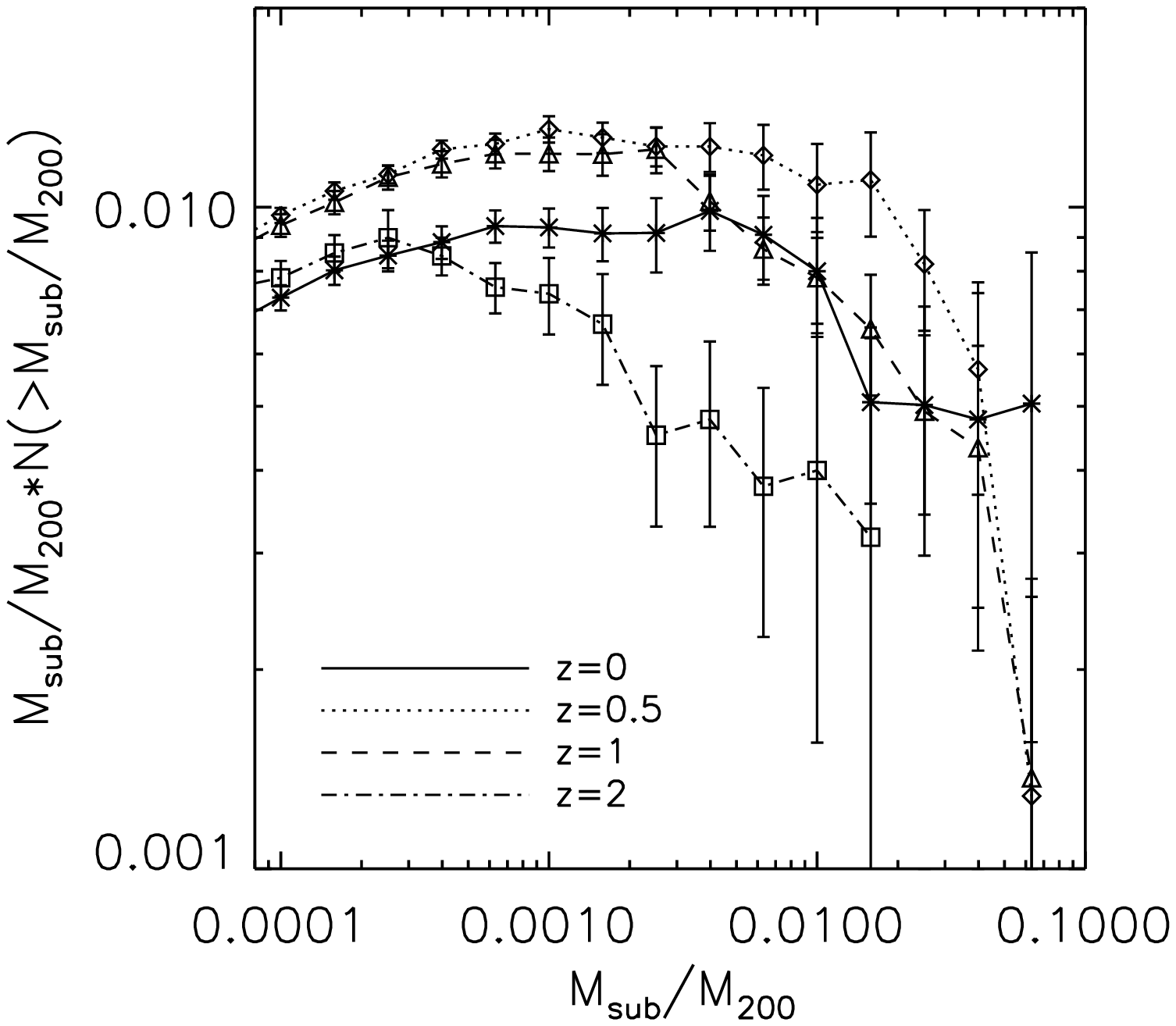} &
\includegraphics[scale=0.50]{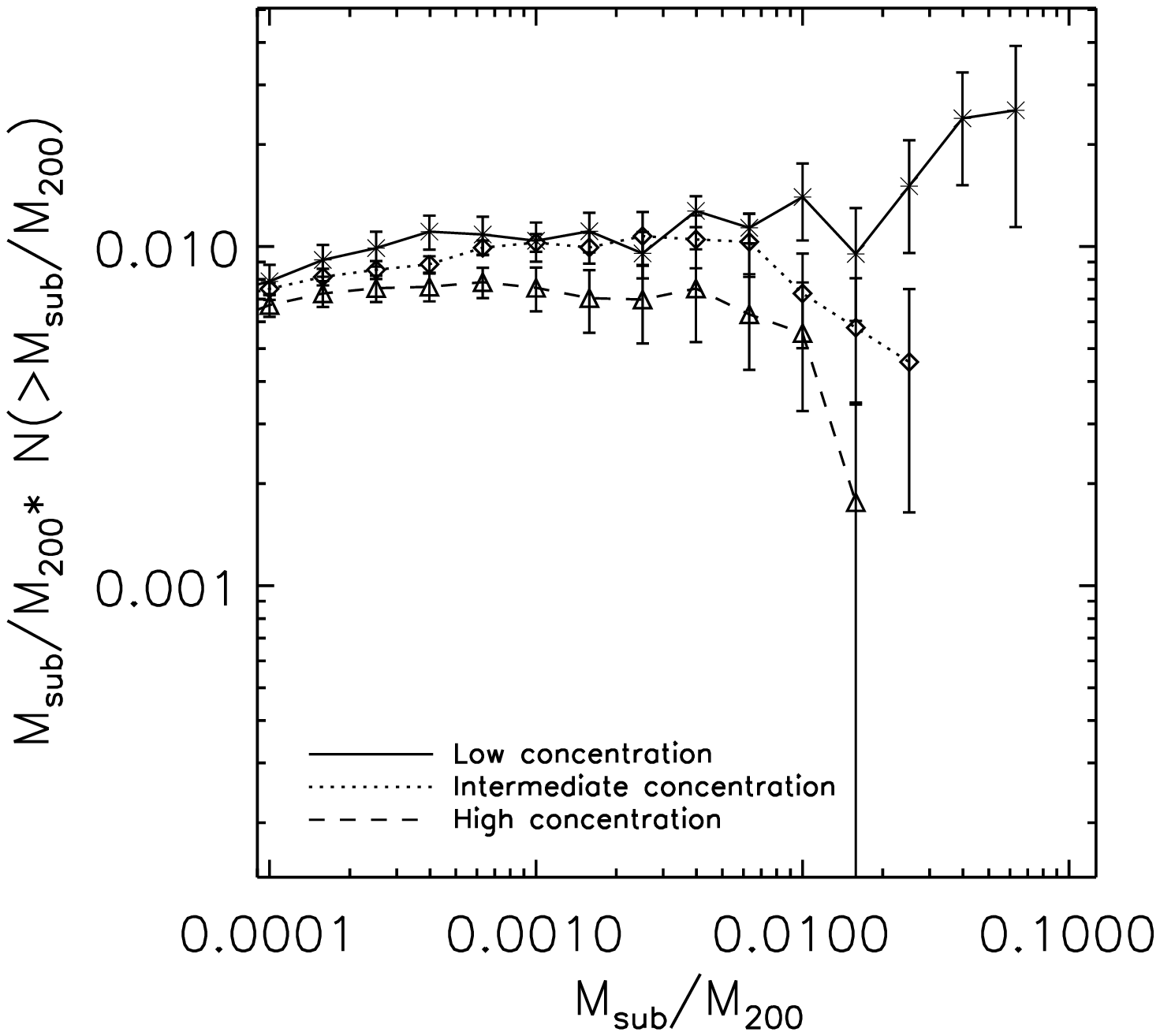} \\
\end{tabular}
\caption{Cumulative mass functions (CMF) in units of rescaled subhalo mass, and
  multiplied by $M_{sub}/M_{200}$ to take out the dominant mass dependence.
  Top and bottom panels are for haloes in the mass range $[1-3]\times
  10^{13}\,\hm M_{\odot}$ and $[1-5]\times 10^{14}\,\hm M_{\odot}$,
  respectively. In the left panel, results are shown for different redshifts
  (solid line for z=0, dotted line for z=0.5, dashed line for z=1 and
  dash-dotted line for z=2). In the right panel, only haloes identified at
  redshift zero have been considered, and they have been split in three bins,
  according to their concentration. Only subhaloes with more than 100 bound
  particles have been used to build these functions.}
\label{fig4}
\end{figure*}
\end{center}

  In Figure \ref{fig2}, we plot the slope of the differential mass function
  obtained by fitting a power law to the mass functions of each sub-sample
  considered in our study. Following \citet{delucia}, we have restricted the
  fit by discarding the most massive (and rarest) substructures (those with
  mass above $10^{12} \, \hm M_{\odot}$ for the samples S1 and S2, and with
  mass above $10^{11.5} \, \hm M_{\odot}$ for the samples S3, S4 and S5). We
  find that, albeit weakly, the slope of the subhalo mass function depends on
  the parent halo mass, and that there is a weak trend for shallower slopes
  with increasing lookback times. The best fit values we measure vary in the
  range between $\sim -0.65$ and $\sim -0.8$, in agreement with results from
  previous studies \citep[e.g.][]{ghigna2,delucia,gao}. When including the most
  massive substructures in the fit, we obtain steeper slopes, ranging from
  $\sim -0.91$ and $\sim -0.86$ at redshift $z=0$, but the trends shown in
  Figure \ref{fig2} are not altered significantly.

  As explained by \citet{gao2}, the dependence of the subhalo mass
  function on halo mass is a consequence of the fact that more massive
  haloes are on average less concentrated and dynamically younger than
  their less massive counterparts. Since the strength of tidal
disruption depends on halo concentration, and since haloes of a given
mass are on average less concentrated at higher redshift, we also
expect that the subhalo mass function depends on
time. Figure~\ref{fig4} shows the cumulative subhalo mass function
(normalized as in \citealt{gao}) at four different redshifts in the
left panels and for different concentrations in the right panels (in
these panels, only haloes identified at redshift zero have been
considered). Top and bottom panels refer to the haloes in the mass
range $[1-3]\times 10^{13}\,\hm M_{\odot}$ and $[1-5]\times
10^{14}\,\hm M_{\odot}$ respectively. We derive the three subsamples
by splitting the range of concentration in order to have the same
number of haloes in each subsample. Results shown in Figure~\ref{fig4}
confirms previous findings by \citet{gao2}, and extend them to larger
parent haloes masses: haloes at higher redshift have significantly
more substructures than those of the same mass at later times. The
figure suggests that there is a significant evolution between $z=0$
and $z\sim 0.5$, but it becomes weaker at higher redshifts. We note
that for the highest redshift considered, the subhalo mass function
does not significantly differ from that found at $z\sim 1$, but we
note that this could be due to poor statistics. \cite{gao2} find a
similar trend for haloes of similar mass. At any given cosmic epoch,
there is a large halo-to-halo scatter which is due, at least in part,
to internal properties of the parent halo like concentration, as shown
in the right panels of Figure~\ref{fig4}. For the ranges of mass shown
in Fig.~\ref{fig4}, low concentration haloes host up to an order of
magnitude more substructures than haloes of the same mass but with
higher concentration. The difference between the different
concentration bins are larger (and significant) for the most massive
substructures.

  In order to verify that the results of our analysis are robust against
  numerical resolution, we have compared the cumulative sub-halo mass function
  obtained for the set of simulated halos presented here to that obtained for
  the same halos simulated at a 10 times lower mass resolution. We find that
  the two distributions agree very well to each other, within the mass range
  accessible to both resolutions. This confirms that both our simulations and
  the procedure of halo identification are numerically converged.

\subsection[]{Radial Distribution of Subhaloes}
\label{sec:raddist}

Previous studies \citep{ghigna2,delucia,nagai,saro} have shown that subhaloes
are `anti-biased' relative to the dark matter in the inner regions of
haloes. No significant trend has been found as a function of the parent halo
mass, with only hints for a steeper profiles of subhaloes in low massive haloes
\citep{delucia}.

\begin{center}
\begin{figure}
\begin{tabular}{c}
\includegraphics[scale=.29]{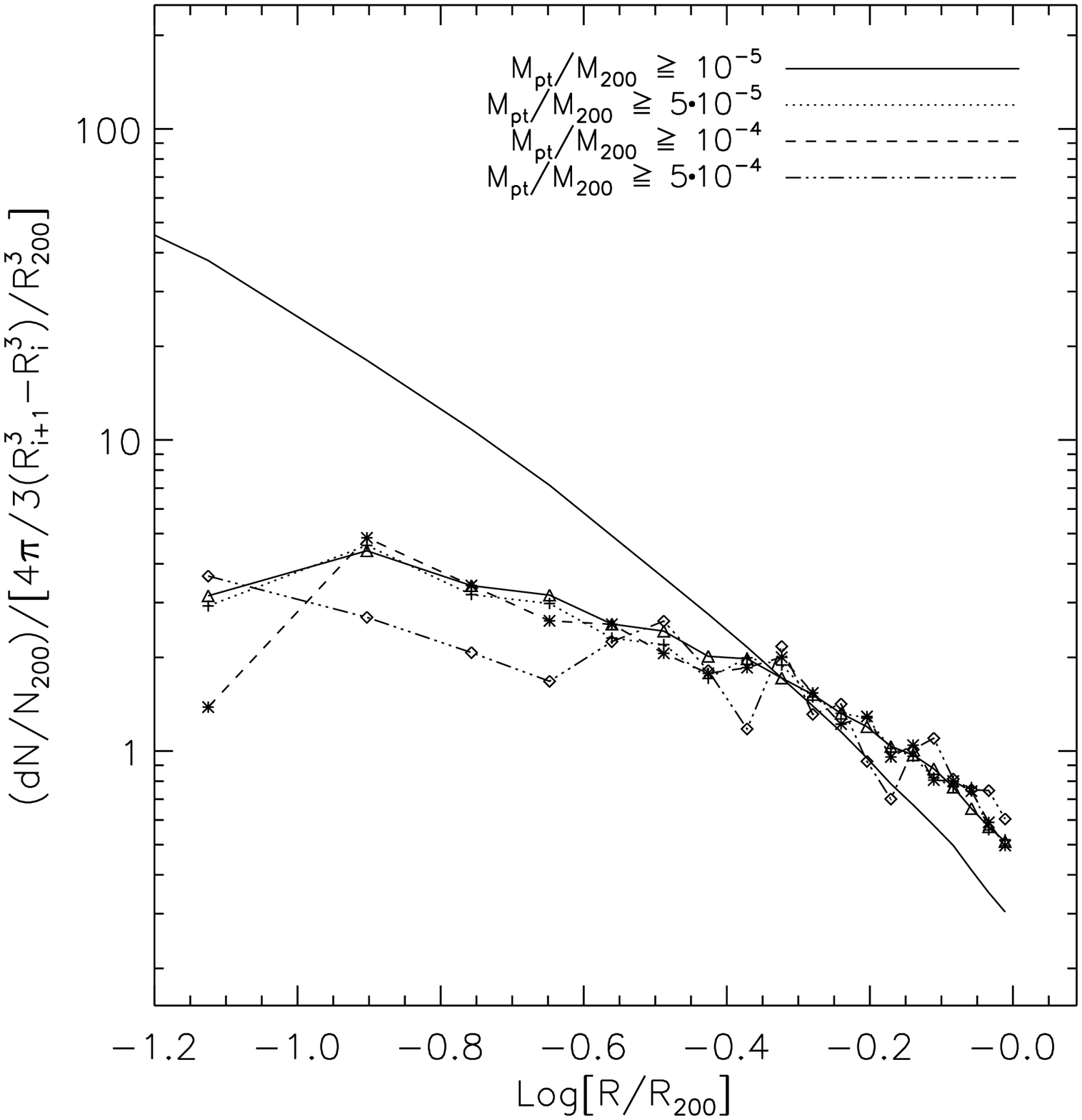} \\
\includegraphics[scale=.29]{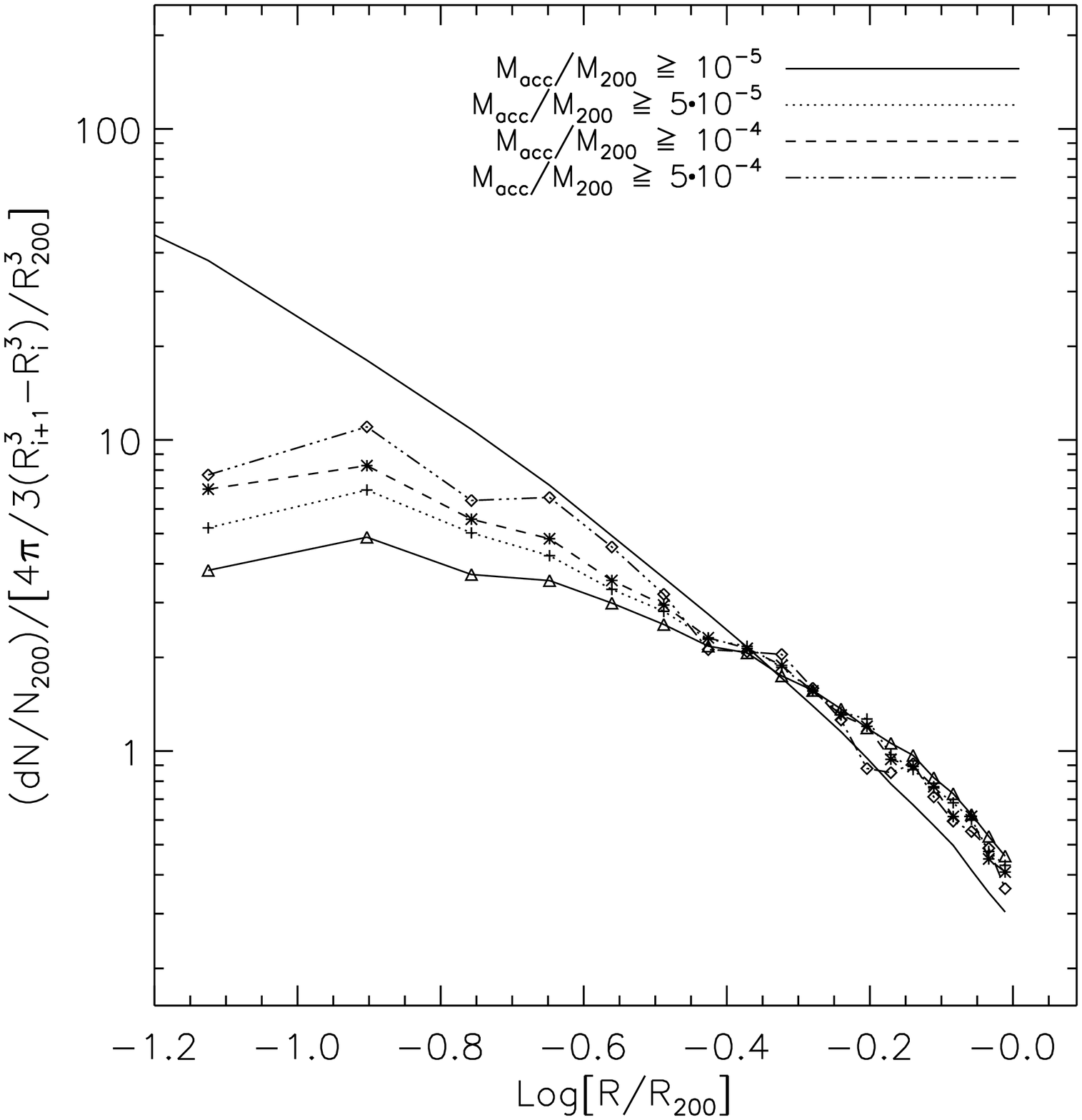} \\
\includegraphics[scale=.29]{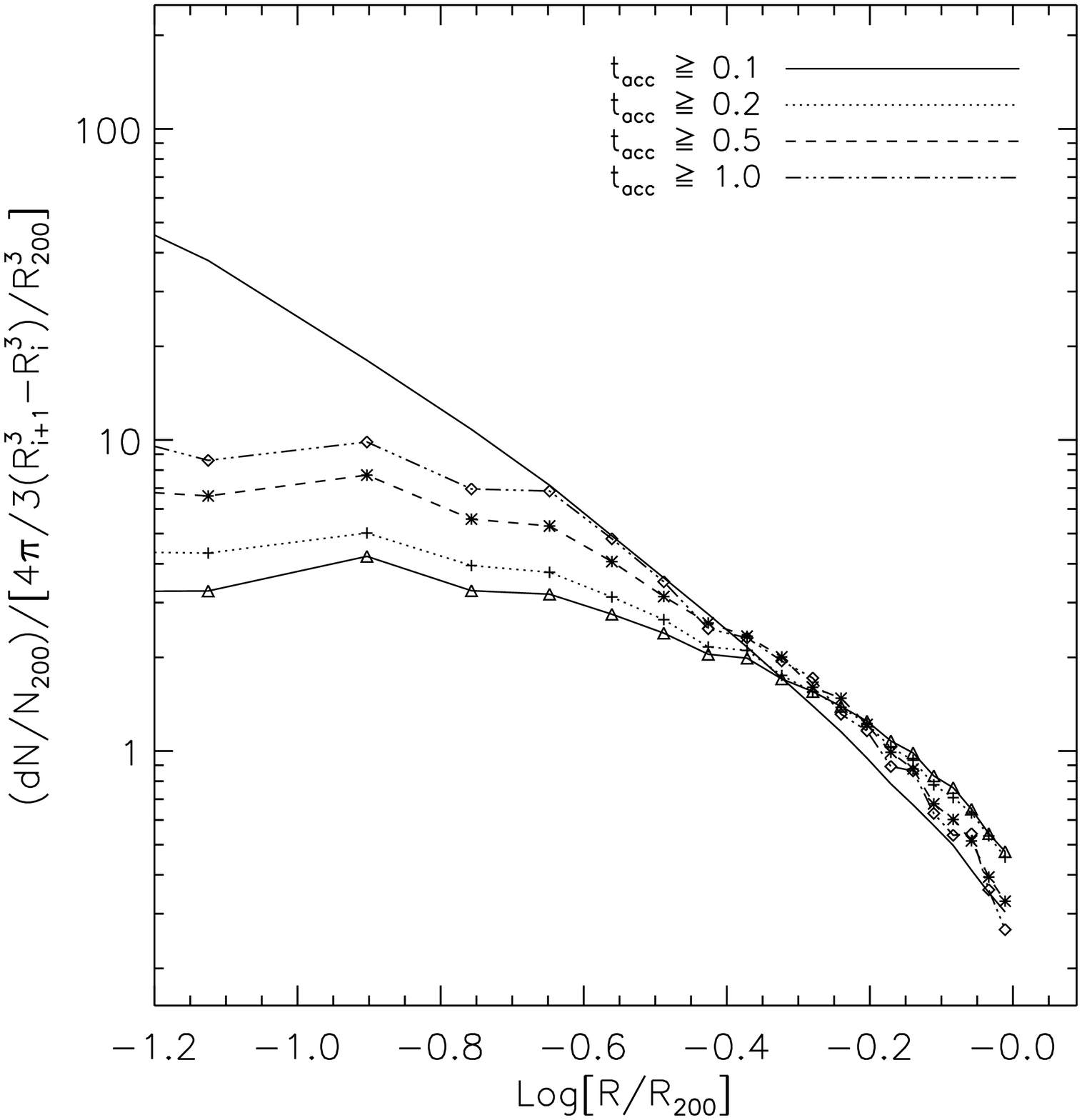} \\
\end{tabular}
\caption{Radial distribution of dark matter substructures
    belonging to haloes of the sample S1. In the top panel, different
    lines correspond to different thresholds in the $M_{sub}/M_{200}$
    ratio, based on the present-day subhalo mass. In the middle panel,
    the subhalo mass at the time of accretion has been considered,
    while in the bottom panel different lines correspond to subhaloes
    accreted at different times.}
\label{fig7b}
\end{figure}
\end{center}

  The analysis of our sample of simulated halos confirms previous
  findings that dark matter subhaloes are anti-biased with respect to
  dark matter, with no dependence on parent halo mass. In fact, there
  is no physical reason to expect such a trend. We note that
  \citet{delucia}, who found hints for such a correlation, used a
  smaller sample of simulated haloes, that were carried out using
  different simulation codes and parameters. In contrast, our
  simulated haloes are all carried out using the same parameters and
  simulation code.

  \cite{nagai} find that the anti-bias is much weaker if subhaloes are
  selected on the basis of the mass they had at the time of accretion
  onto their parent halo. We confirm their results in
  Figure~\ref{fig7b}, where we show the radial distribution of
  substructures in our sample S1 (the most massive haloes in our
  simulation set). The top panel of Figure~\ref{fig7b} shows the
  radial distribution of substructures selected on the basis of their
  present day mass, while in the middle panel the mass of the
  substructure at the time of accretion (defined as the last time the
  halo was identified as a central halo, see below) has been used. The
  figure shows that, in this case, selecting progressively more
  massive substructures reduces the anti-bias between subhaloes and
  dark matter.  The bottom panel of Figure~\ref{fig7b} shows that the
  same is obtained by discarding substructures that are accreted
  recently. The two selections tend to pick up haloes that suffered a
  stronger dynamical friction (i.e. haloes that were more massive at
  the time of accretion) or that suffered of dynamical friction for a
  longer time (haloes that were accreted earlier). As a consequence,
  both selections tend to preferentially discard subhalos at larger
  radii, thus bringing the radial distribution of subhaloes closer to
  that measured for dark matter.

\begin{figure}
\begin{center}
\includegraphics[scale=.43]{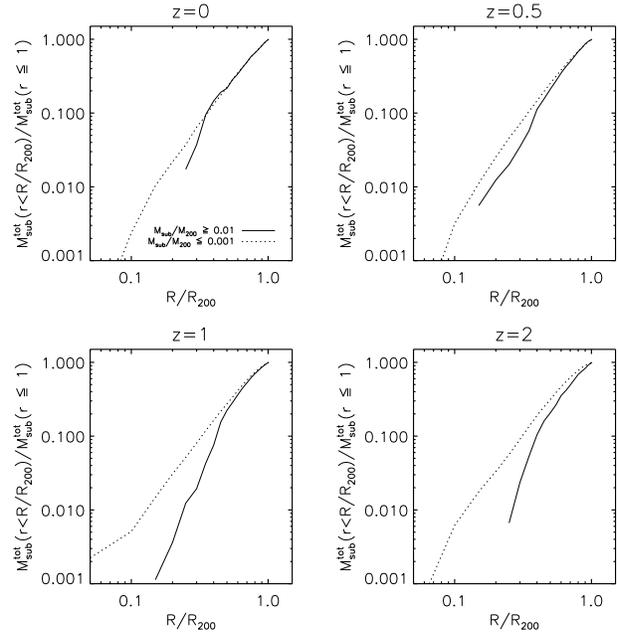}
\caption{Cumulative radial distributions for subhaloes with $M_{sub}/M_{200} >
  0.01$ (solid line) and $M_{sub}/M_{200} < 0.001$ (dotted line) from all
  samples, at different redshifts. On the y-axis, we plot the total mass in
  subhaloes within a given distance from the centre, normalized to the total
  mass in subhaloes within $R_{200}$, for each subhaloes population.}
\label{fig9}
\end{center}
\end{figure}

As shown above (see right panels of Figure \ref{fig5}), most of the
substructure mass is located at the cluster outskirts. \citet{delucia} showed
that this distribution is dependent on the subhalo mass, with the most massive
substructures being located at larger distances from the cluster centre with
respect to less massive substructures. In particular, \cite{delucia} split
their subhalo population in two subsamples by choosing a rather arbitrary mass
ratio between the subhalo mass and the parent halo mass (they chose the value
0.01 for this ratio). Our simulations exhibit the same trends, but we find that
this can be more or less `significant' depending on the particular threshold
adopted to split the sample. In Figure~\ref{fig9}, we show the radial
distribution of substructures with $M_{sub}/M_{200}> 0.01$ (solid lines) and
$M_{sub}/M_{200}< 0.001$ (dashed lines). Our trends are weaker than those found
by \citet{delucia} at redshift zero, when the same division is adopted. We
note, however, that these trends are dominated by the most massive substructure
and are, therefore, significantly affected by low number
statistics. Figure~\ref{fig9} also shows that the mass segregation becomes more
important at increasing redshift.

Considering that haloes of a given mass are less centrally concentrated and
dynamically younger than their counterparts at later redshift, the trend found
can be explained as follows: the `younger' haloes have massive subhaloes
preferentially in their outer regions because stripping has not had enough time
to strip their outer material and eventually disrupt them. In more dynamically
evolved clusters (those at present time), stripping has had more time to
operate and to wash out any difference between the two distributions.  In this
picture, the balance between dynamical friction and stripping on one hand, and
the accretion of new subhaloes on the other hand is such that the latter effect
is dominating over the former. This is in agreement with the results shown
above for the evolution of the cumulative mass function of substructures, whose
normalization increases with increasing redshift.

We stress that in Figure~\ref{fig9} we are considering subhaloes of different
mass at the time they are identified. As discussed in Section~\ref{sec:intro},
this cannot be simply related to the mass and/or luminosity of the galaxies. So
the trend shown in Figure~\ref{fig9} cannot simply be related a different
spatial distribution for galaxies in different luminosity bins, as done for
example in \citet[][ see their figure 8]{lin}.

\section[]{Evolution of Substructures}
\label{sec:subevol}

In this section, we study the evolution of substructures as a function of time,
focusing in particular on their mass accretion histories and merger histories.
In order to obtain these information, we have constructed merger histories for
all self-bound haloes in our simulations, following the method adopted in
\citet{springel3} and the improvements described in \citet{delucia2}.  

Briefly, the merger tree is constructed by identifying a unique {\it
  descendant} for each substructure. For each subhalo, we find all haloes that
contain its particles in the following snapshot, and then count the particles
by giving higher weight to those that are more tightly bound to the halo under
consideration. The halo that contains the largest (weighted) number of its
particles is selected as descendant. Next, all the pointers to the progenitors
are constructed. By default, the most massive progenitor at each node of the
tree is selected as the {\it main progenitor}. \citet{delucia2} noted that this
can lead to ambiguous selections when, for example, there are two subhaloes of
similar mass. In order to avoid occasional failures in the merger tree
construction algorithm, they modified the definition of the main progenitor by
selecting the branch that accounts for most of the mass of the final system,
for the longest time. We have applied this modification to our merger trees. In
this section, we consider only substructures that contain at least 100 bound
particles, and in a few cases, we use particular mass ranges to ease the
comparison with the literature.

In this section we will also study if the accretion and merger history of
substructures depend on the environment, that we will approximate using the
parent halo mass. It is worth stressing, however, that our haloes provide
likely a biased sample for this analysis. In fact, excluding the most massive
sample and some haloes that belong to the sample S2, all the other haloes 
reside in the regions surrounding the most massive
haloes, which might not represent the `typical' environment for halo in the same
mass range. 

\subsection[]{Mass Accretion History}
\label{sec:history}

In this section, we use the merger trees constructed for our cluster sample to
study the mass accretion histories of subhaloes of different mass and residing
in different environments. Several previous studies \citep{gao,delucia,warnick}
have pointed out that once haloes are accreted onto larger systems (i.e. they
become substructures), their mass is significantly reduced by tidal
stripping. The longer the substructure spends in a more massive halo, the
larger is the destructive effect of tidal stripping. Previous studies have
found that the efficiency of tidal stripping is largely independent of the
parent halo mass \citep{delucia,gao}.

We re-address these issues using all substructures residing within the virial
radius of our haloes, and with mass larger than $10^{10}\,\hm M_{\odot}$ at
redshift $z=0$ (in our simulations, these substructures contain at least 100
particles). By walking their merger trees, following the main progenitor
branch, we construct the mass accretion history (MAH) for all of these
subhaloes, and record the accretion time ($z_{accr}$) as the last time the halo
is a central halo, i.e. before it is accreted onto a larger structure and
becomes a proper subhalo. Our final sample includes 39005 haloes, that we split
in two bins of different mass by using either their present day mass or their
mass at the accretion time. We end up with 33576 haloes with mass larger than
$10^{11}\,\hm M_{\odot}$ at present (25344 when using the mass at the accretion
time), and 5429 haloes with mass lower than the adopted threshold (13661 if the
accretion mass is used). In order to analyse the environmental dependence of
the mass accretion history, we consider separately subhaloes residing in our S5
and S1 samples (these correspond to our lowest and largest parent halo mass,
respectively).

\begin{center}
\begin{figure*}
\begin{tabular}{cc}
\includegraphics[scale=0.53]{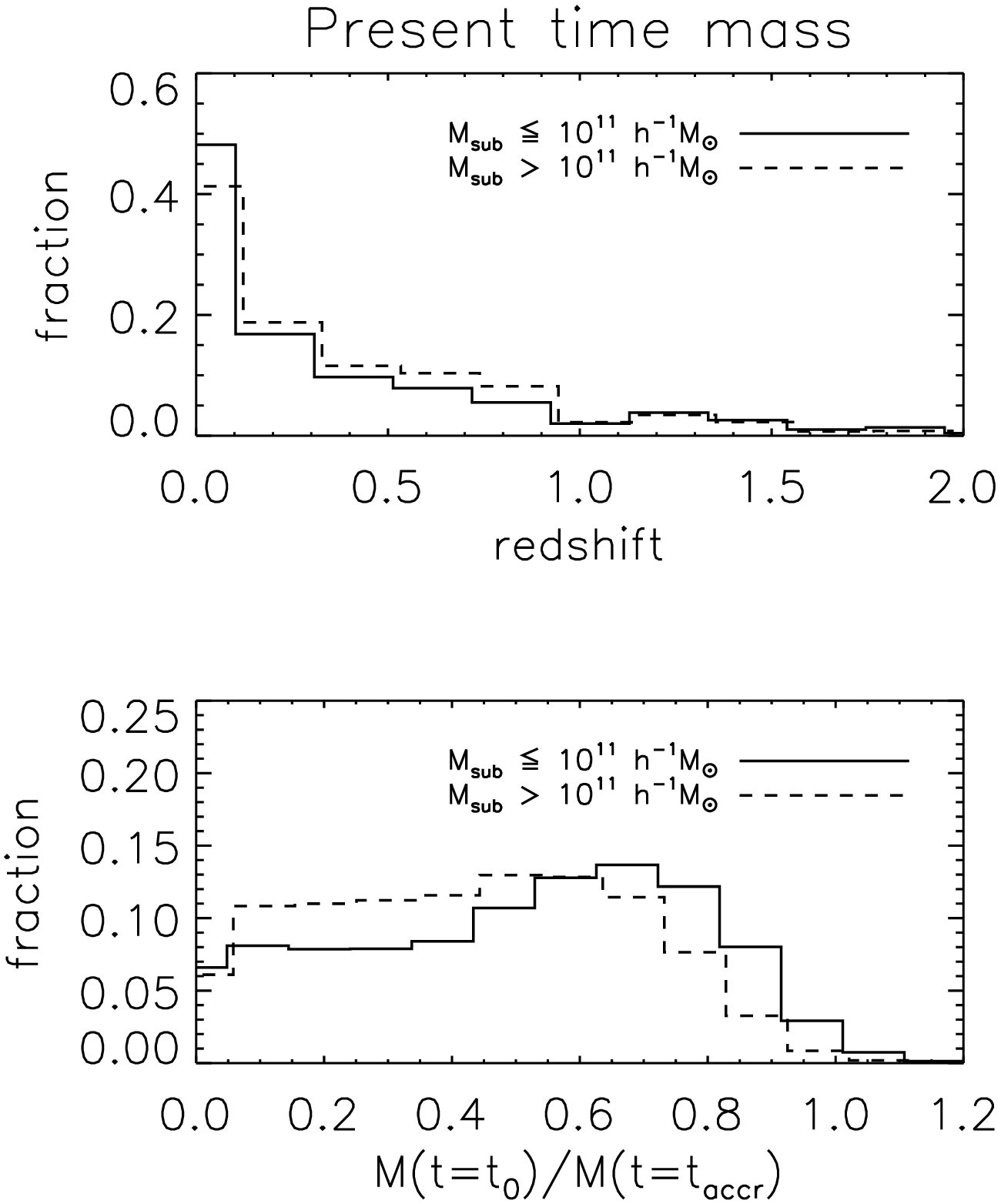} &
\includegraphics[scale=0.53]{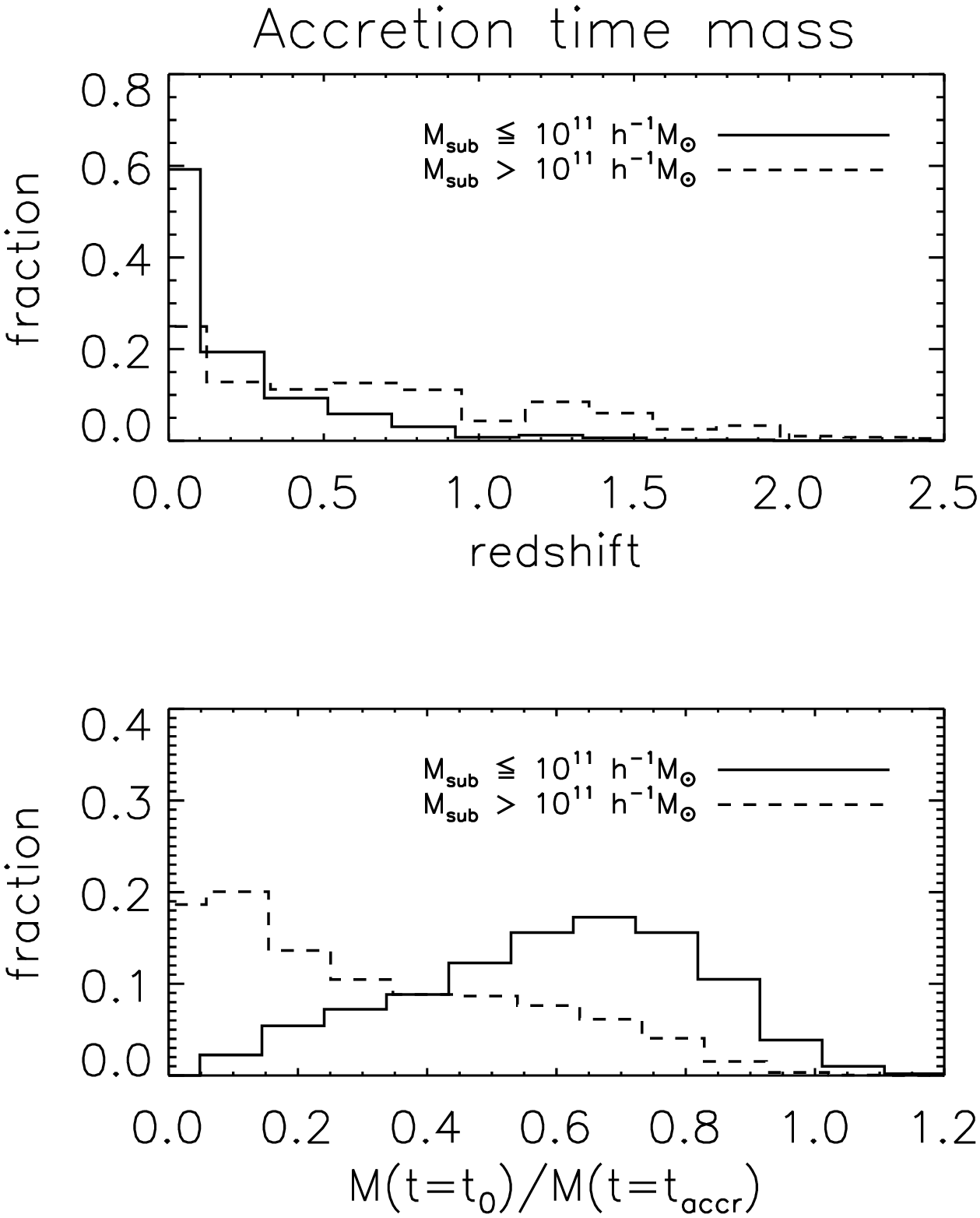} \\
\end{tabular}
\caption{Left panels: distribution of the accretion times (top panel) and of
  the fraction of mass loss since accretion (bottom panel) for subhaloes of
  different mass at present time (different linestyles, as indicated in the
  legend). Right panels show the same distributions but for subhaloes split
  according to their mass at the accretion time.}
\label{zaccdist2}
\end{figure*}
\end{center}

The top panels in Figure~\ref{zaccdist2} show the distribution of the accretion
times for the two mass bins considered. Left and right panels correspond to a
splitting in mass done on the basis of the present day mass and of the mass at
accretion, respectively. When considering the present day mass (left panel),
the differences between the two distributions are small, with only a slightly
lower fraction of more massive substructures being accreted very late, and a
slightly larger fraction of substructures in the same mass range being accreted
between $z\sim 0.1$ and $z\sim 1$. A larger difference between the two
distribution can be seen when considering the mass at the time of accretion
(right panel). Substructures that are less massive at the time of accretion
have been accreted on average later than their more massive counterparts. In
particular, about 90 per cent of the substructures in the least massive bin
considered have been accreted below redshift 0.5, while only 50 per cent of the
most massive substructures have been accreted over the same redshift range. The
distribution obtained for the most massive substructures is broader, extending
up to redshift $\sim 2$. This is largely a {\it selection effect}, due to the
fact that we are only considering substructures that are still present at
$z=0$. Once accreted onto larger systems, substructures are strongly affected
by tidal stripping so that, among those that were accreted at early times, only
the most massive ones will still retain enough bound particles at present to
enter our samples. The less massive substructures that were accreted at early
times, have been stripped below the resolution of our simulations and therefore
do not show up in the solid histogram that is shown in the top right panel of
Figure~\ref{zaccdist2}. 

\begin{center}
\begin{figure}
\includegraphics[scale=0.55]{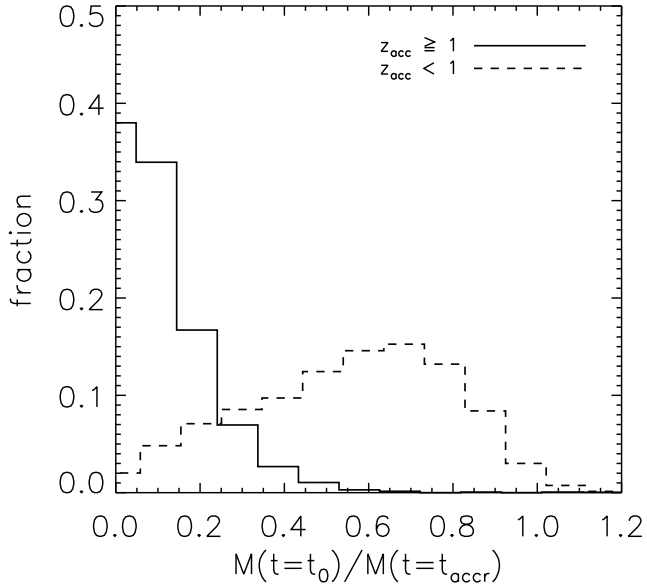}
\caption{Distribution of mass loss (ratio between the present day mass and the
  mass at accretion) for two different accretion ranges: solid line for
  $z_{accr}\geq 1$ and dotted line for $z_{accr}<1$.}
\label{zaccdist}
\end{figure}
\end{center}

The bottom panels of Figure~\ref{zaccdist2} show the distribution of the ratios
between present day mass and mass at accretion for subhaloes of different
present day mass (left panel) and for different mass at accretion (right
panel). Less massive subhaloes, which were accreted on average more recently,
lose on average smaller fractions of their mass compared to more massive
subhaloes for which the distribution is skewed to higher values. The difference
between these distributions becomes more evident when one split the samples
according to the mass at the time of accretion, as shown in the right panel. As
explained above, however, this is affected by the fact that many of the least
massive substructure will be stripped below the resolution of the simulation at
$z=0$.  We have repeated the analysis done in Fig. \ref{zaccdist2} for
subhaloes in each of the five samples used in our study, and we found there is
no significant dependency on the environment.

Fig. \ref{zaccdist} shows that, as expected, substructures accreted earlier
suffered significantly more stripping than substructures that were accreted at
later times. In particular, about 90 per cent of subhaloes accreted at redshift
larger than 1 have been stripped by more than 80 per cent of their mass at
accretion. For haloes that have been accreted at redshift lower than 1, the
distribution is much broader, it peaks at $\sim 0.6$ (i.e. about 40 per cent of
the mass has been stripped for about 20 per cent of these haloes) but has a
long tail to much lower values.  Similarly to Fig. \ref{zaccdist2}, we also
tried to split this plot for different parent halo masses, without finding any
significant trend with the environment.

\begin{center}
\begin{figure*}
\begin{tabular}{cc}
\includegraphics[scale=0.56]{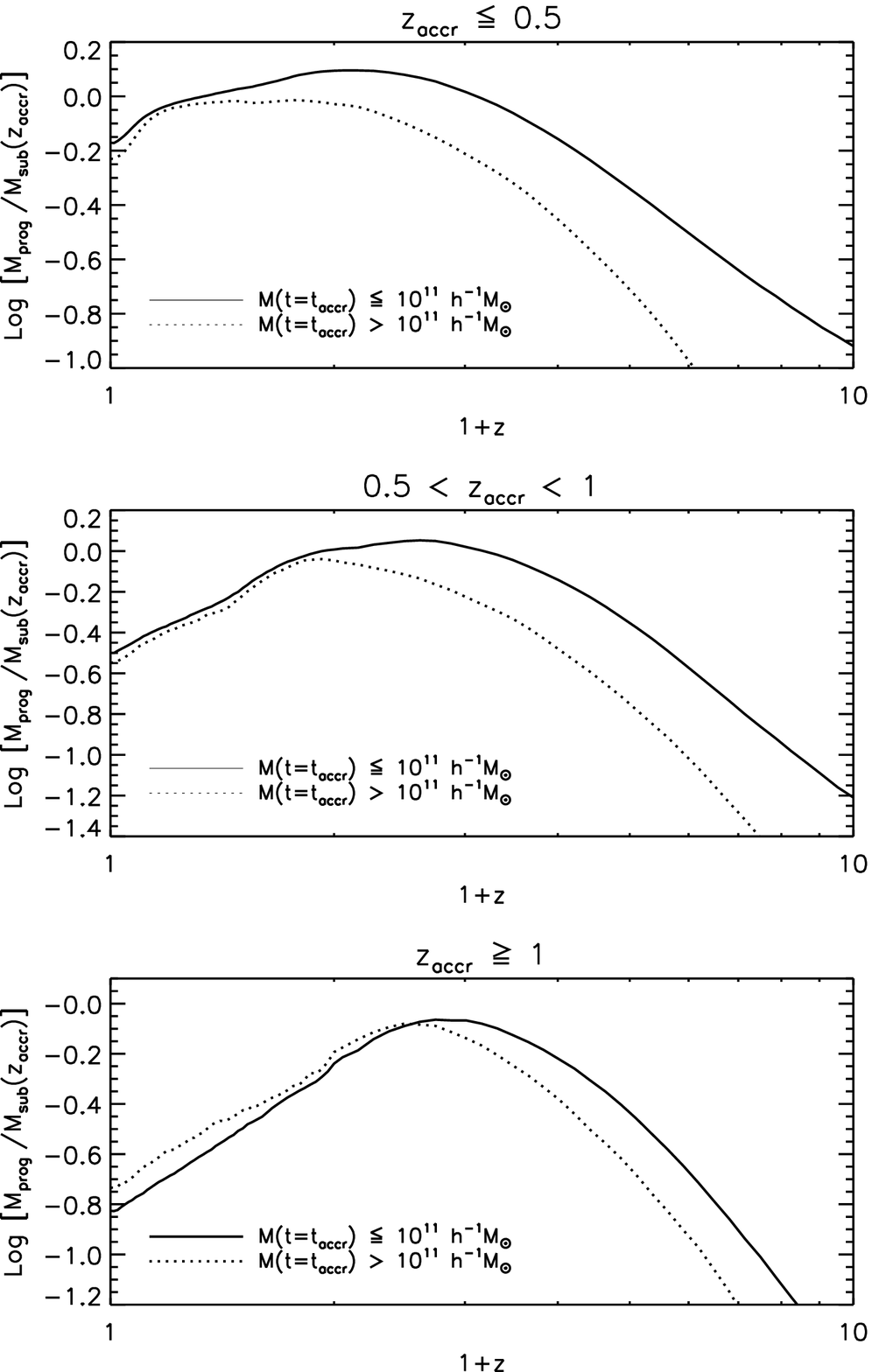} &
\includegraphics[scale=0.56]{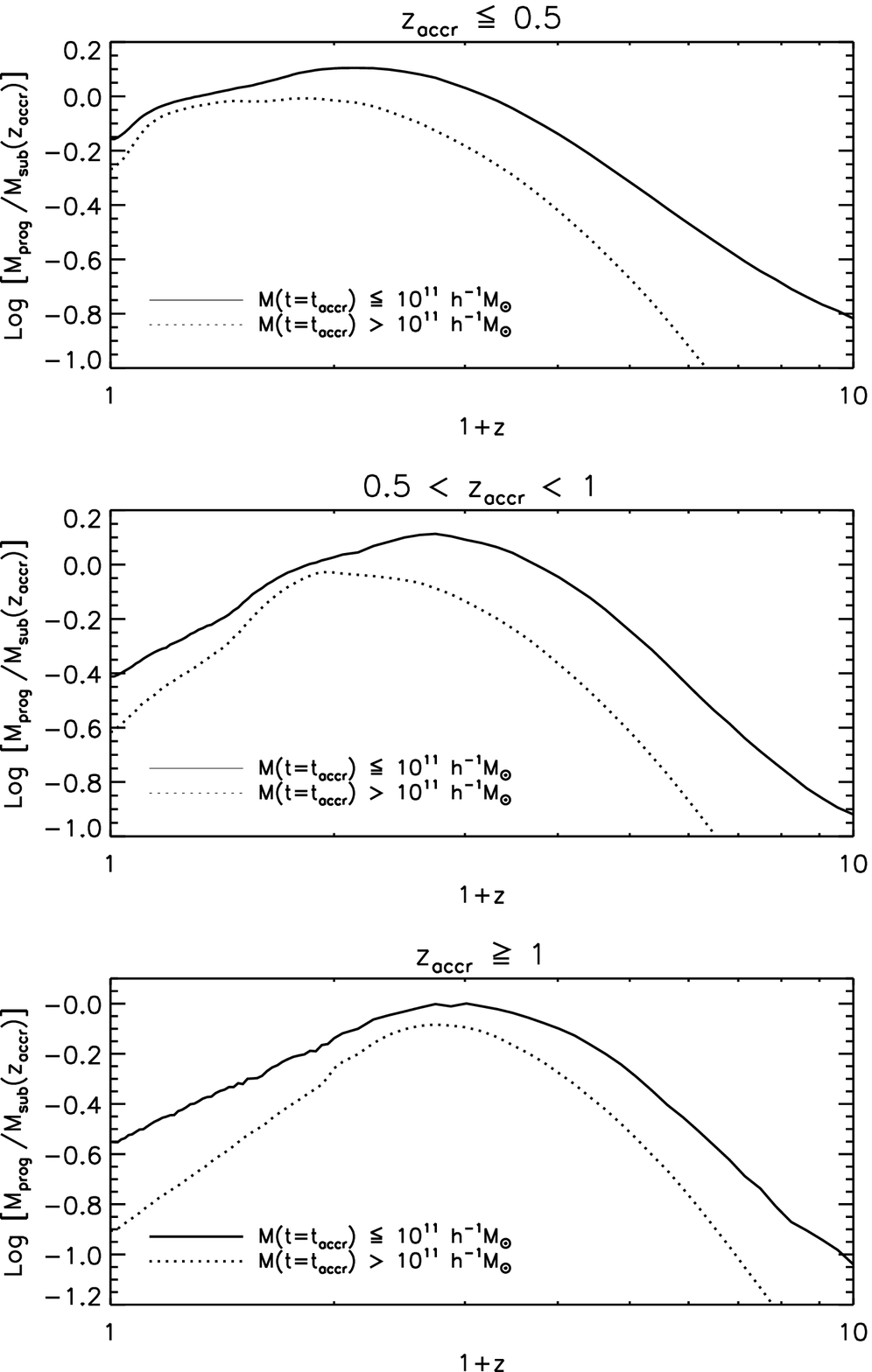} \\
\end{tabular}
\caption{Average mass accretion history for three ranges of accretion times. In
  the left panels, substructures are split according to their present day mass,
  while in the right panels they are split according to their mass at the time
  of accretion.}
\label{mah}
\end{figure*}
\end{center}

Fig. \ref{mah} shows the MAHs of subhaloes accreted at different times. It
shows results when subhaloes are split according to their present day mass
(left panels), and the mass at accretion (right panels).  As shown in previous
studies, the longer the halo is a substructure, the larger is its stripped
mass. When substructures are split according to their present day mass, the
influence of tidal stripping does not appear to depend strongly on the
substructure mass. In contrast, if the mass at the accretion time is
considered, in a given range of accretion times, haloes that are more massive
lose a larger fraction of their mass with respect to their less massive
counterparts. This is due to the fact that more massive haloes sink more
rapidly towards the centre because of dynamical friction, and therefore suffer
a more significant stripping due to tidal interactions with the parent
halo. Once again, this entails the fact that luminosity must correlate stronger
with the subhalo mass computed at the time of accretion, i.e. before stripping
had time to operate.

\begin{figure}
\begin{center}
\includegraphics[scale=.55]{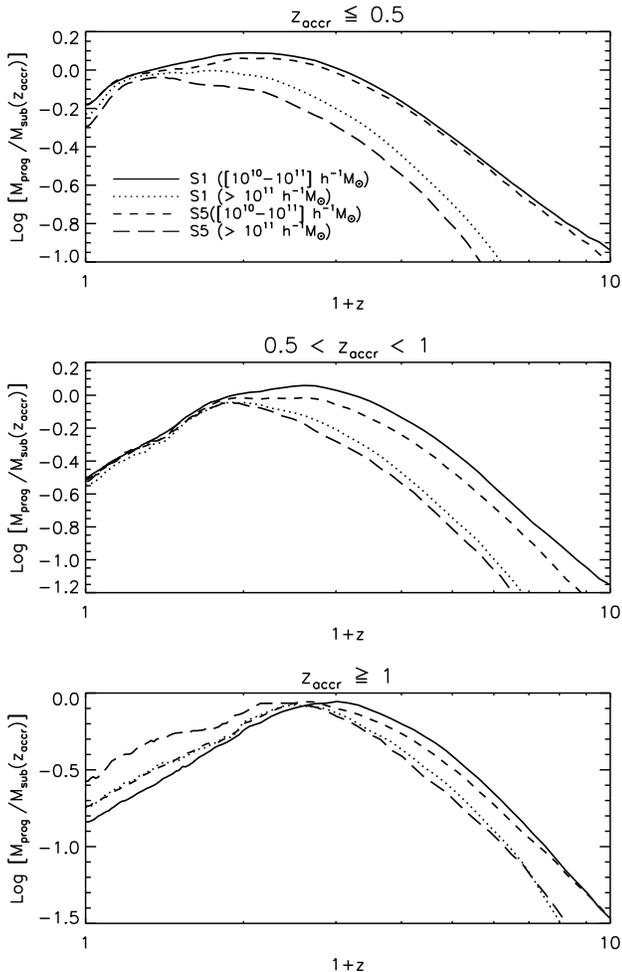}
\caption{Average mass accretion history for subhaloes in three different ranges
  of accretion time, as a function of environment. Dashed and long-dashed lines
  show the MAH for subhaloes in the sample S5 with mass in the range
  $[10^{10}-10^{11}]\,\hm M_{\odot}$ and larger than $10^{11}\,\hm M_{\odot}$,
  respectively. Solid and dotted lines show the MAH for subhaloes in the same
  mass ranges but for the sample S1.}
\label{mah2}
\end{center}
\end{figure}

In Fig. \ref{mah2} we plot the mean MAHs for subhaloes in the two mass bins
considered and for two different `environments', parametrized as the mass of
the parent halo. In particular, we consider the samples S5 and S1 (i.e. the
least and the most massive haloes used in our study). Dashed and long-dashed
lines show the MAHs for subhaloes in the sample S5 with mass in the range
$[10^{10}-10^{11}]\,\hm M_{\odot}$ and larger than $10^{11}\,\hm M_{\odot}$,
respectively. Solid and dotted lines show the MAH for subhaloes in the same
mass ranges but for the sample S1. Here we consider the present day subhalo
mass.  Computing the same plot by adopting the subhalo mass at the time of
accretion does not alter the results. We find that the environment does not
significantly influence the mass accretion history of substructures. In the
bottom panel, the long dashed line (corresponding to substructures more massive
than $10^{11}\,\hm M_{\odot}$ in the sample S5) is likely affected by low
number statistics. In the same panel a small difference can be seen for the
less massive substructures that appear to be less stripped in the sample S5
than in S1 (compare dashed and solid lines). The difference, however, is not
large, but this might be affected by the fact that our haloes all reside in the
regions surrounding very massive clusters. 

\subsection{Merging Rate}
\label{sec:mergrate}

In recent years, a large body of observational evidence has been collected that
demonstrates that galaxy interactions and mergers play an important role in
galaxy evolution. In particular, numerical simulations have shown that major
mergers between two spiral galaxies of comparable mass can completely destroy
the stellar disk and leave a kinematically hot remnant with structural and
kinematical properties similar to those of elliptical galaxies \citep*[][ and
  references therein]{mvw}.  Minor mergers and rapid repeated encounters with
other galaxies residing in the same halo (\textit{harassment};
\citealt{moore2}; \citealt{moore3}) can induce disk instabilities and/or the
formation of a stellar bar, each of which affects the morphology of galaxies falling onto
clusters. As galaxy mergers are driven by mergers of the parent dark matter
haloes, it is interesting to analyse in more detail the merger statistics of
dark matter substructures.

The mass accretion history discussed in the previous section does not
distinguish between merger events (of different mass ratios) and accretion of
`diffuse material'. In order to address this issue, and in particular to study
the merger rates of dark matter substructures, we have taken advantage of the
merger trees constructed for our samples. We have selected all subhaloes with
mass larger than $10^{12}\,\hm M_{\odot}$ at redshift zero, and have followed
them back in time by tracing their main progenitor branch, and recording all
merger events with other structures. In particular, we take into account only
mergers with objects of mass larger than $10^{10}\,\hm M_{\odot}$, and mass
ratios larger than $5:1$. We note that both these values are computed at the
time the halo is for the last time central (the mass of the main progenitor at
the time of accretion is considered to compute the mass ratio).

\begin{figure}
\begin{center}
\includegraphics[scale=.53]{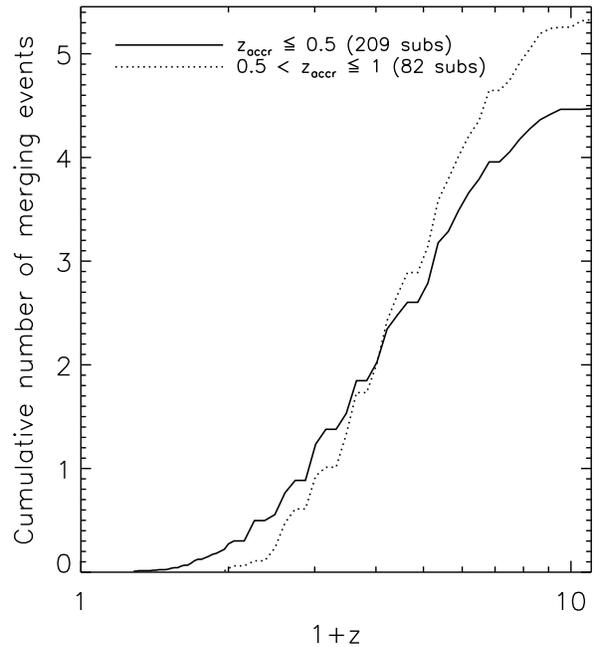}
\caption{Mean number of major mergers as a function of redshift, for subhaloes
  in two different ranges of accretion time. We take into account only
  subhaloes with mass $M \geq 10^{12}\,\hm M_{\odot}$ at redshift $z=0$ and
  merger events that include systems with mass $M \geq 10^{10}\,\hm
  M_{\odot}$.}
\label{merg}
\end{center}
\end{figure}

\begin{figure}
\begin{center}
\includegraphics[scale=.53]{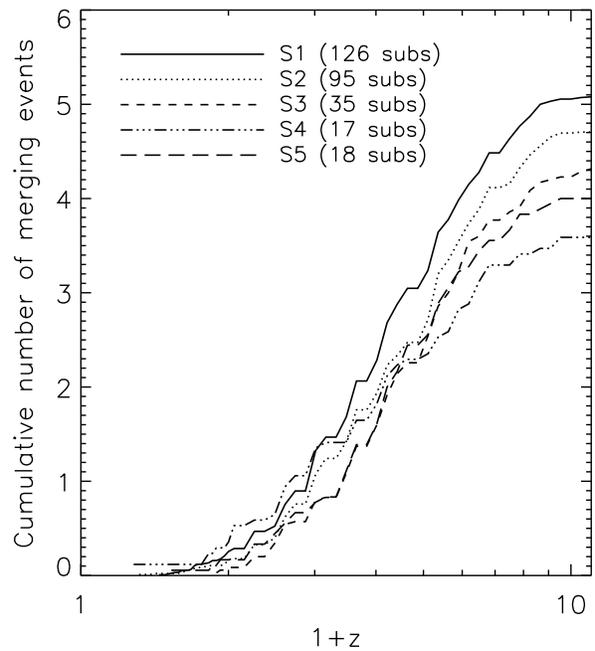}
\caption{Mean number of major mergers as a function of redshift, for subhaloes
  in different environments, quantified as the mass of their parent halo.  As
  in Fig. \ref{merg} we take into account only subhaloes with mass $M \geq
  10^{12}\,\hm M_{\odot}$ at redshift $z=0$ and merger events that include
  systems with mass $M \geq 10^{10}\,\hm M_{\odot}$.}
\label{mergenv}
\end{center}
\end{figure}

Fig. \ref{merg} shows the merging rate for all subhaloes that satisfy the above
conditions. We consider in this plot only objects that experienced at least one
merger event. The solid line shows the mean number of mergers for subhaloes
that were accreted at $z<0.5$, while the dotted line shows the resulting merger
rate for objects accreted between $0.5< z \leq 1$. The figure shows that in
both cases, the slope of the lines become shallower close to the accretion
time, i.e. mergers between substructures are suppressed because of the large
velocity dispersion of the parent haloes. Interestingly, haloes that were
accreted earlier experience, on average, one more major merger than haloes
accreted at later times. 

We repeat the same analysis looking at the merging rate as a function of
environment. Fig. \ref{mergenv} shows the cumulative number of mergers for
subhaloes in our five samples. The mean number of mergers increases as a
function of the parent halo mass, although subhaloes in the sample S4
experience on average fewer mergers than subhaloes in the sample S5. This is
not surprising since subhaloes in the surroundings of more massive haloes have
a larger probability to merge with other structures. 

  \section{Discussion and conclusions}
\label{sec:concl}

We have used a large set of high-resolution simulated haloes to
analyse the statistics of subhaloes in dark matter haloes, and their
dependency as a function of the parent halo mass and physical
properties of the parent halo. While some of the results
  discussed in this study confirm results from previous studies, it is
  the first time that a systematic analysis of the properties and
  evolution of dark matter substructures is carried out using a large
  simulation set carried out using the same cosmological parameters
  and simulation code. Our main results can be summarized as follows:

\begin{itemize}

\item[(i)] More massive haloes contain increasing fractions of mass in
  subhaloes.  This does not exceed $\sim 10$ per cent of the total mass, in
  agreement with previous studies. There is, however, a very large halo-to-halo
  scatter that can be partially explained by a range of halo physical
  properties, e.g. the concentration. Indeed, in more concentrated haloes
  substructures suffer of a stronger tidal stripping so that they are
  characterized by lower fractions of mass in substructures.

\item[(ii)] We find that the subhalo mass function depends weakly on the parent
  halo mass and on redshift. This can be explained by considering that haloes
  of larger mass are less concentrated and dynamically younger than their less
  massive counterparts, and that haloes of a given mass are on average less
  concentrated at higher redshift. Our findings confirm results from previous
  studies \citep{gao2}, and extend them to larger halo masses.

\item[(iii)] As shown in previous work \citep[e.g.][]{ghigna,delucia},
  subhaloes are anti-biased with respect to the dark matter in the
  inner regions of haloes. The anti-bias is considerably reduced
    once subhaloes are selected on the basis of their mass at the time
    of accretion, or neglecting those that were accreted at later
    times. We also find that the spatial distribution of subhaloes
  does not depend significantly on halo mass, as suggested in previous
  work by \citet{delucia}. The most massive substructures are located
  at the outskirts of haloes and this mass segregation is more
  important at higher redshift.

\item [(iv)] Once accreted onto larger systems, haloes are strongly affected by
  tidal stripping. The strength of this stripping appears to depend on the mass
  of the accreting substructures: those that are more massive at the time of
  accretion tend to be stripped by larger fractions of their initial mass. 

\item [(v)] Mergers between substructures are rare events. Following the merger
  trees of substructures, however, we find that they have suffered in the past
  about 4-5 important (mass ratio 1:5) mergers. As expected, the number of
  mergers experienced depends on the environment: subhaloes in more massive
  systems have experienced more mergers than those of similar mass residing in
  less massive haloes. 
\end{itemize}

Dark matter substructures mark the sites where luminous satellites are expected
to be found, so their evolution and properties do provide important information
on the galaxy population that forms in hierarchical models. As discussed in
previous studies, however, because of the strong tidal stripping suffered by
haloes falling onto larger structures, it is not possible to simply correlate
the population of subhaloes identified at a given cosmic epoch to that of the
corresponding galaxies. The galaxy luminosity/stellar mass is expected to be
more strongly related to the mass of the substructure at the time of {\it
  infall} and, depending on the resolution of the simulations, there might be a
significant fraction of the galaxy population that cannot be traced with dark
matter substructures because they have been stripped below the resolution limit
of the simulation (the `orphan' galaxies - see for example \citealt{wang}). 

Nevertheless, our results do provide indications about the properties of the
galaxy populations predicted by hierarchical models. Tidal stripping is largely
independent of the environment (we have parametrized this as the parent halo
mass), while the accretion rates of new subhaloes increases at increasing
redshift. The nearly invariance of the subhalo mass function results from the
balance between these two physical processes. If the amount of dark matter
substructures is tracing the fraction of recently infallen galaxies, the
fraction of star forming galaxies is expected to increase with increasing
redshift (the `Butcher-Oemler' effect, \citealt{butcher},
\citealt{kauffmann}). In addition, our findings suggest that stronger mass
segregation should be found with increasing redshift.

There is a large halo-to-halo scatter that can be only partially explained by a
wide range of physical properties. This is expected to translate into a large
scatter in e.g. the fraction of passive galaxies for haloes of the same mass,
with more concentrated haloes hosting larger fraction of red/passive galaxies.
Finally, there is an obvious merger bias that is expected to translate into a
different morphological mix for haloes of different mass. In future work, we
plan to carry out a more direct comparison with observational data at different
cosmic times, by applying detailed semi-analytic model to the merger trees
extracted from our simulations.    

\section*{Acknowledgements}
We thank the anonymous referee for constructive comments that helped
  improving the presentation of the results. EC and GDL acknowledge financial
support from the European Research Council under the European Community's
Seventh Framework Programme (FP7/2007-2013)/ERC grant agreement n. 202781. This
work has been supported by the PRIN-INAF 2009 Grant ``Towards an Italian
Network for Computational Cosmology'' and by the PD51 INFN grant. Simulations
have been carried out at the CINECA National Supercomputing Centre, with CPU
time allocated through an ISCRA project and an agreement between CINECA and
University of Trieste.  We acknowledge partial support by the European
Commissions FP7 Marie Curie Initial Training Network CosmoComp
(PITN-GA-2009-238356).  We thank A. Bonafede and K. Dolag for their help with
the initial conditions of the simulations used in this study.

\bsp

\label{lastpage}


\bibliographystyle{mn2e}
\bibliography{biblio}

\end{document}